\documentclass[prl,showpacs,byrevtex,floatfix,twocolumn,superscriptaddress]{revtex4}
\usepackage{amssymb,graphicx,afterpage}
\begin{document}
\title{\boldmath Magnetic order induced crystal symmetry lowering in ACr$_2$O$_4$ ferrimagnetic spinels \unboldmath}
%
%
%
\author{S. Bord\'acs}
\affiliation{Department of Physics, Budapest University of
Technology and Economics and Condensed Matter Research Group of the Hungarian Academy of Sciences, 1111 Budapest, Hungary} \affiliation{Multiferroics Project, ERATO, Japan Science and Technology Agency (JST), Japan c/o Department of Applied Physics, The University of Tokyo, Tokyo 113-8656, Japan}
\author{D. Varjas}
\affiliation{Department of Physics, Budapest University of
Technology and Economics and Condensed Matter Research Group of the Hungarian Academy of Sciences, 1111 Budapest, Hungary}
\author{I. K\'ezsm\'arki}
\affiliation{Department of Physics, Budapest University of
Technology and Economics and Condensed Matter Research Group of the Hungarian Academy of Sciences, 1111 Budapest, Hungary} \affiliation{Multiferroics Project, ERATO, Japan Science and Technology Agency (JST), Japan c/o Department of Applied Physics, The University of Tokyo, Tokyo 113-8656, Japan} \affiliation{Department of Applied
Physics, University of Tokyo, Tokyo 113-8656, Japan}
\author{G. Mih\'aly}
\affiliation{Department of Physics, Budapest University of
Technology and Economics and Condensed Matter Research Group of the Hungarian Academy of Sciences, 1111 Budapest, Hungary}
\author{L. Baldassarre}
\affiliation{Experimentalphysik 2, Universit\"at Augsburg, D-86135 Augsburg, Germany}
\author{A. Abouelsayed}
\affiliation{Experimentalphysik 2, Universit\"at Augsburg, D-86135 Augsburg, Germany}
\author{C.A. Kuntscher}
\affiliation{Experimentalphysik 2, Universit\"at Augsburg, D-86135 Augsburg, Germany}
\author{K. Ohgushi}
\affiliation{Institute for Solid State Physics, University of Tokyo, Kashiwanoha, Kashiwa, Chiba 277-8581, Japan}
\author{Y. Tokura}
\affiliation{Multiferroics Project, ERATO, Japan Science and Technology Agency (JST), Japan c/o Department of Applied Physics, The University of Tokyo, Tokyo 113-8656, Japan}
\affiliation{Department of Applied Physics, University of Tokyo,
Tokyo 113-8656, Japan}
\date{\today}
%
%
\pacs{\ }
\begin{abstract}
We demonstrate that the onset of complex spin orders in ACr$_2$O$_4$ spinels with magnetic A$=$Co, Fe and Cu ions lowers the lattice symmetry. This is clearly indicated by the emergence of anisotropic lattice dynamics -- as evidenced by the pronounced phonon splittings -- even when experiments probing static distortions fail. We show that the crystal symmetry in the magnetic phase is reduced from tetragonal to orthorhombic for FeCr$_2$O$_4$ and CuCr$_2$O$_4$ with Jahn-Teller active A-site ions. The conical spin structure in FeCr$_2$O$_4$ is also manifested in the phonon frequencies. In contrast, the multiferroic CoCr$_2$O$_4$ with no orbital degrees of freedom remains nearly cubic in its ground state.
\end{abstract}
\maketitle

The coupling between electronic and lattice degrees of freedom is an essential ingredient of correlated electron physics and inevitably manifested in many of the related phenomena. The representative examples, when the two degrees of freedom indeed show cooperative orderings, range from the highlighted family of ferroelectrics \cite{Kimura07} through materials exhibiting cooperative Jahn-Teller effect \cite{Millis96} to systems with spin Peierls \cite{Hirota94}, charge and/or spin density wave ground state \cite{Grunerbook}. Perhaps the most delicate pattern, where all the orbital, charge, spin and lattice degrees of freedom are interlocked, is realized in manganites being the mother state for the colossal magnetoresistance effect \cite{Dagotto02}.
Not less common and fundamental are the various manifestations of the electron-phonon coupling in the charge transport such as the conventional superconductivity, the polaron motion in bad metals \cite{Ronnow06}, the Fano asymmetry of optical phonon modes \cite{Kezsmarki06} or the appearance of phonon resonances in the differential conductivity of quantum wires \cite{Agrait02}, etc.

Recently, the interaction between spin and lattice degrees of freedom has attracted much interest either in the context of magneto-elastic effects \cite{Lee08} or as a possible path to get rid off the ground state degeneracy in frustrated magnets \cite{Tchernyshyov02,Penc04}. It is of general expectation that the onset of a symmetry breaking \emph{magnetic order can in principle change the underlying lattice} via the spin-lattice coupling. A classic example is the trigonal distortion in the N\'eel state of MnO, NiO with rocksalt structure where the exchange interactions serve as the primary driving force of the transition and the lowering of the lattice symmetry from cubic to rhombohedral is a subsidiary consequence \cite{Roth58,Massida99}.

On the other hand, in highly frustrated antiferromagnets a secondary energy scale, though generally much weaker than that of the spin-spin interaction, can play an essential role in the development of long range magnetic order. Such phenomenon was observed in several ACr$_2$O$_4$ chromite spinels with nonmagnetic A-site ions and termed as spin Jahn-Teller effect \cite{Tchernyshyov02,Penc04} -- being analogous to the ordinary Jahn-Teller effect when the orbital degeneracy of an ion is lifted by the lowering of the crystal symmetry. In these spinels, the frustration of J$_{Cr-Cr}$ antiferromagnetic nearest neighbor interaction on the corner sharing network of Cr$^{3+}$ tetrahedra, the so-called pyrochlore structure, is eliminated by a tetragonal distortion of the lattice. Consequently the spin degeneracy is lifted and a unique antiferromagnetic ground state appears. Beyond the static structural distortion, the anisotropy of the lattice dynamics, i.e. considerable splitting of some infrared active phonon modes, was found upon the antiferromagnetic transition and has attracted interest as a consequence of the strong spin-phonon coupling which derives from the spatial dependence of the exchange interactions \cite{Tchernyshyov02,Fennie06,Rudolf07}.

\begin{figure}[t!]
\includegraphics[width=2in]{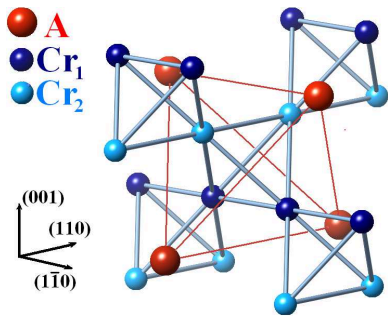}
\caption{(Color online) Representative part of the spinel structure with Cr$^{3+}$ ions (smaller balls) forming a pyrochlore and A$^{2+}$ ions (bigger balls) a diamond lattice. ACr$_2$O$_4$ chromites with magnetic ion on the A-site (Mn, Fe, Co, Ni and Cu) -- irrespective of the details of the spin order -- contain three magnetic sublattices A, Cr$_1$ and Cr$_2$ in their ferrimagnetic phase \cite{Ohgushi08}. The net magnetization generally points along one of the main cubic axes with the Cr$_1$ and Cr$_2$ chains running in the normal plane being also perpendicular to each other.}
\label{fig1}
\end{figure}

In this Letter, we study representatives from another class of spinel compounds where the A$^{2+}$ ions forming diamond lattice are also magnetic, namely CoCr$_2$O$_4$, FeCr$_2$O$_4$ and CuCr$_2$O$_4$. If the J$_{A-Cr}$ coupling between the A-site and the Cr$^{3+}$ S$=3/2$ spins becomes large enough, it helps to overcome the frustration of the pyrochlore lattice and may result in a complex noncollinear order with three magnetic sublattices \cite{Menyuk07} -- A, Cr$_1$ and Cr$_2$ as displayed in Fig.~\ref{fig1}. Conical spin structure was indeed observed in CoCr$_2$O$_4$ and FeCr$_2$O$_4$ below T$_S$$=$$27$\,K and $35$\,K, respectively \cite{Shirane64,Tomiyasu04,Yamasaki06}. Though ferrimagnetic order is already present above the conical phase -- below T$_C$$=93$\,K and $97$\,K for A$=$Co and Fe -- the spin components perpendicular to the spontaneous magnetization are still disordered \cite{Tomiyasu04}. On the other hand, CuCr$_2$O$_4$ becomes a canted antiferromagnet at T$_C$$=$$152$\,K with again three magnetic sublattices \cite{Prince57}.

CoCr$_2$O$_4$ has no orbital degree of freedom since the ground
state of both the Co$^{2+}$ (d$^7$) and Cr$^{3+}$ (d$^3$) ions is an orbital singlet
due to the tetrahedral and octahedral crystal field, respectively. Accordingly,
this compound is reported to retain its cubic spinel structure down to low temperatures with the space group Fd$\overline{3}$m \cite{Yamasaki06,Menyuk07}. In FeCr$_2$O$_4$ and CuCr$_2$O$_4$ the A-site ions are Jahn-Teller active with orbital degeneracy at high temperatures which is lifted
by a tetragonal distortion at T$_{JT}$=135\,K and T$_{JT}$=850\,K respectively
\cite{Tanaka66,Kennedy08}. The factor group
analysis yields four infrared active T$_{1u}$ modes in a cubic spinel.
When the symmetry is reduced to tetragonal (corresponding to D$_{4h}$ point group) we should find that triple degenerate T$_{1u}$'s split into a singlet and a doublet, T$_{1u}$=A$_{2u}$$\bigoplus$E$_u$ \cite{note}. For further lowering of the symmetry to orthorhombic (D$_{2h}$ point group), the degeneracy of the modes is fully lifted according to T$_{1u}$=A$_{2u}$$\bigoplus$B$_{1u}$$\bigoplus$B$_{2u}$ and 12 optical vibrations are expected in the infrared spectrum.

Our primary goal was to follow the signatures of the magnetic transitions on the lattice dynamics by infrared spectroscopy. We succeeded in observing the splitting of several modes pointing out that the \emph{spin ordering does reduce the symmetry of the lattice}, as well.

Single crystals of CoCr$_2$O$_4$ and FeCr$_2$O$_4$ were grown by
chemical vapor transport while CuCr$_2$O$_4$ was prepared
by flux decomposition method \cite{Ohgushi08}. Reflectivity spectra, both polarized and unpolarized, was measured with microscopes on the as-grown surfaces  ([111] for A$=$Co, Fe and [010] for A$=$Cu) between T=10-300\,K.  Although our focus is on the lattice vibrations in the far-infrared region we have carried out the experiments over a broad energy range ($\omega$=100-320000\,cm$^{-1}$ and $\omega$=100-40000\,cm$^{-1}$ at room and low temperatures, respectively) to facilitate the proper Kramers-Kronig transformation.

Low-temperature optical conductivity spectra of the ACr$_2$O$_4$ spinels are shown in Fig.~2 in the range of the optical vibrations. Only four phonon modes are discernable in the spectra of CoCr$_2$O$_4$. In contrast, each T$_{1u}$ mode of CuCr$_2$O$_4$ is splitted and the lowest energy mode in FeCr$_2$O$_4$ also has a multi-peak structure indicating symmetry lowering in their originally cubic spinel structure. For a rigorous analysis we have fitted the optical conductivity as a sum of Lorentz oscillators,
\begin{equation}
\sigma(\omega)=\frac{\omega}{4\pi i}\left(\epsilon_{\infty}-1+\sum_i\frac{S_i}{\omega_i^2-\omega^2-i\omega\gamma_i}\right),
\end{equation}
\begin{figure}[t!]
\includegraphics[width=3in]{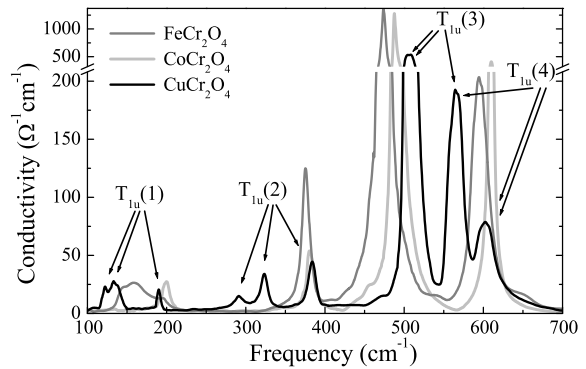}
\caption{Real part of the optical conductivity spectrum for FeCr$_2$O$_4$, CoCr$_2$O$_4$, CuCr$_2$O$_4$ at T=10\,K in the range of infrared active phonons. Arrows indicate the splitting of the three fold degenerate T$_{1u}$ modes in CuCr$_2$O$_4$.}
\label{fig2}
\end{figure}
\begin{figure*}[t!]
\includegraphics[height=2.3in]{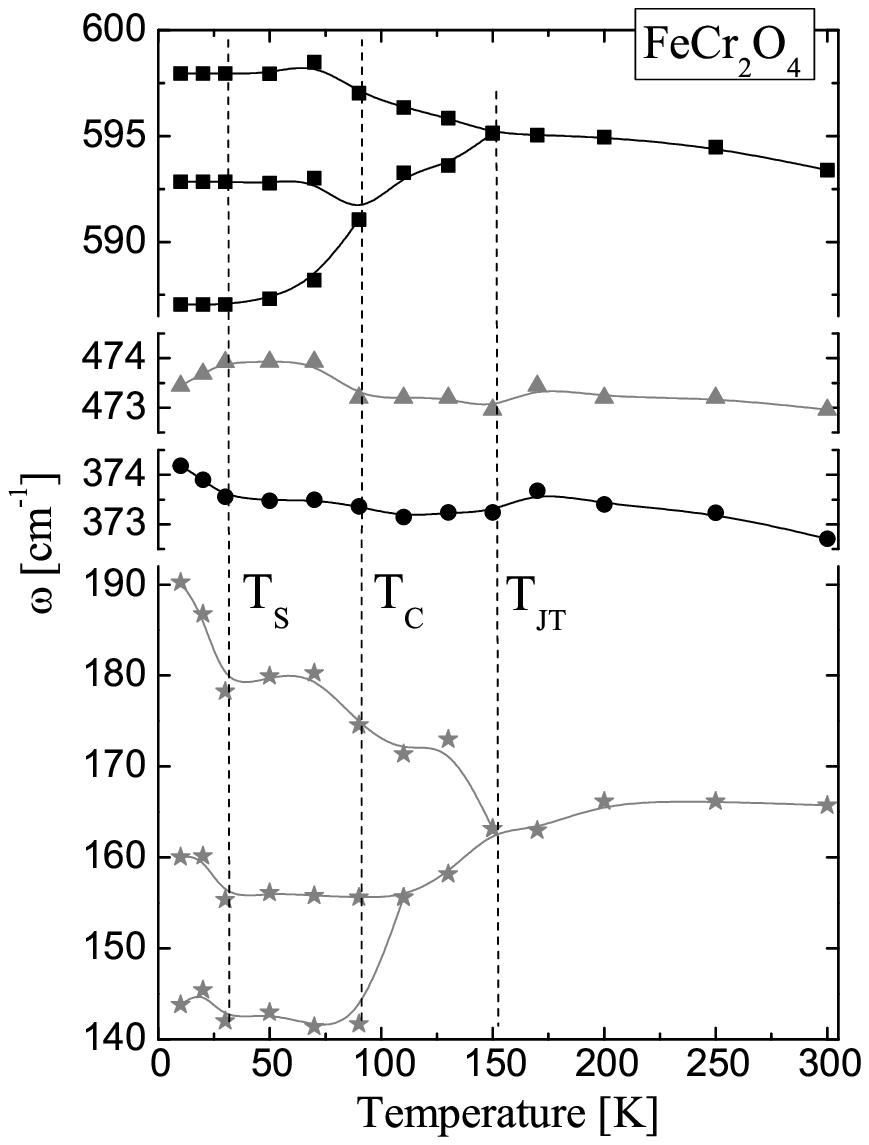}
\hspace{-0.05in}
\includegraphics[height=2.3in]{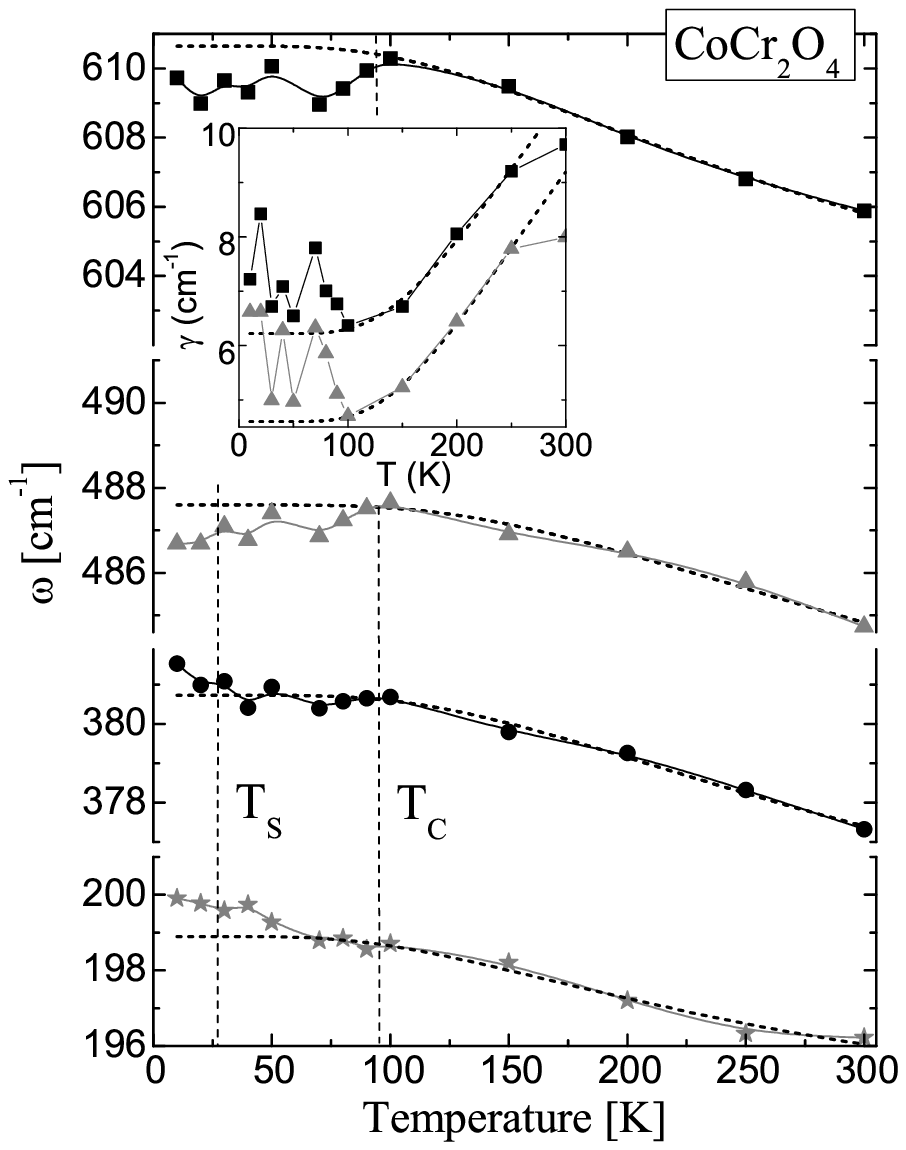}
\hspace{-0.05in}
\includegraphics[height=2.3in]{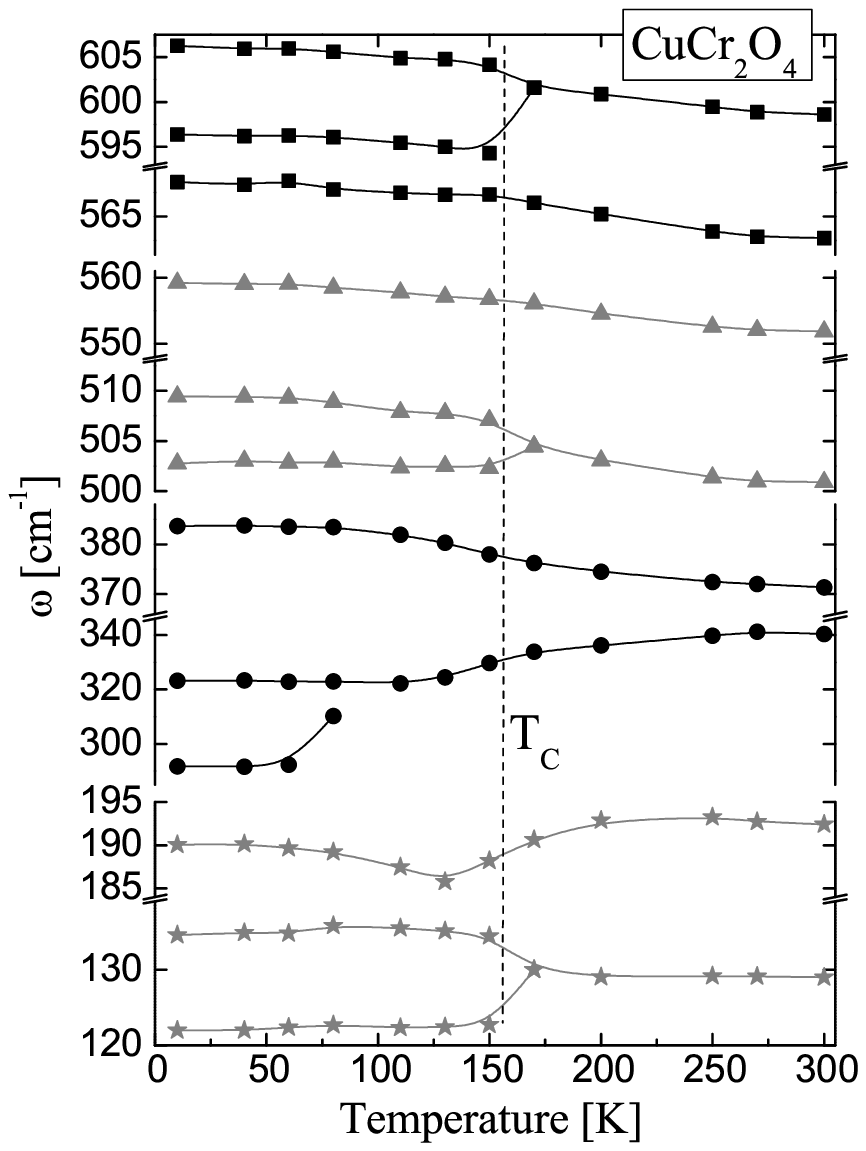}
\caption{Temperature dependence of the phonon frequencies in FeCr$_2$O$_4$, CoCr$_2$O$_4$ and CuCr$_2$O$_4$. T$_{JT}$, T$_C$ and T$_S$ represent transitions to the tetragonal, ferrimagnetic and conical spin state, respectively. Modes originating from the same T$_{1u}$ are labeled by common symbols. The inset shows the damping rate of two modes in CoCr$_2$O$_4$. Dashed lines fitted to the high-temperature part of the $\omega_i(T)$ and $\gamma_i(T)$ curves indicate the conventional behavior due to phonon-phonon anharmonicity \cite{Rudolf07}.}
\label{fig3}
\end{figure*}
where S$_i$, $\omega_i$ and $\gamma_i$ are the oscillator strength, the frequency and the damping rate of the modes and $\epsilon_{\infty}$ is the high-frequency dielectric constant. In CuCr$_2$O$_4$, we observed the large splitting ($\Delta\omega$$\approx$50\,cm$^{-1}$) of each T$_{1u}$ mode into a doublet and a singlet at room temperature as the consequence of the cooperative Jahn-Teller distortion below T$_{JT}$=850\,K. The lifting of degeneracy for the lowest and highest energy mode in FeCr$_2$O$_4$ is similarly observed below T=150\,K in accordance with T$_{JT}$=135\,K. The temperature dependence of the phonon frequencies can be followed in Fig.~3. Besides the splitting of two modes in FeCr$_2$O$_4$, the other phonon lines show anomalous shift upon the Jahn-Teller transition.

We found that the modes, whose degeneracy is partly lifted by the tetragonal distortion, exhibit further splitting at the ferrimagnetic transition. This magnetically induced splitting is well-resolved for each of these phonon lines being as large as $\Delta\omega/\omega_0$$\approx$$10$\% for the lower energy ones. Furthermore, in FeCr$_2$O$_4$ the ferroelectric transition is accompanied by a sudden hardening of the phonon modes originating from T$_{1u}(1)$. The sequence of the mode-splittings upon T$_{JT}$ and T$_C$ is exemplified in Fig.~4 by the temperature dependence of the second mode in CuCr$_2$O$_4$. The polarization dependence of the spectra shown over the same region proves that the structural distortion is a static, long range one and both the structural and magnetic domain sizes exceed the spot size of $\sim50$\,$\mu$m used in our measurements.

Though we could not resolve any splitting of the phonon lines in CoCr$_2$O$_4$, the damping rate $\gamma$ of the two higher energy ones seemingly shows a remarkable increase upon the magnetic ordering as followed in the inset of Fig.~3. (The same is valid for the third mode in FeCr$_2$O$_4$ at T$_{JT}$ and T$_C$, as well.) This is in strong contrast with the conventional behavior characteristic of all the other modes where the damping rate either follows a monotonic decrease in accordance with the freezing out of phonon-phonon scattering or even exhibits a sudden drop upon the transition due to the vanishing of spin-phonon scattering in the magnetically ordered phase. Therefore, this anomalous enhancement is likely the signature of tiny splittings ($\Delta\omega$$\thickapprox$1-2\,cm$^{-1}$) mostly masked by the large damping rates.

The complete lifting of phonon degeneracy in CuCr$_2$O$_4$ and FeCr$_2$O$_4$ according to T$_{1u}$$\rightarrow$A$_{2u}$$\bigoplus$B$_{1u}$$\bigoplus$B$_{2u}$ gives an evidence for the magnetically induced lowering of the lattice symmetry at least to orthorhombic. The corresponding D$_{2h}$ point group is fully compatible with the crystal structure shown in Fig.~1 with the following assumptions: Cr$_1$ and Cr$_2$ sites are necessarily distinguishable in sense of magnetic quantities and spin-lattice coupling can transmit this non-equivalence to primarily spin-independent (or spin integrated) properties such as crystal structure, lattice dynamics, dielectric response, etc. On this basis we conclude that the lattice symmetry in the magnetic phase of CoCr$_2$O$_4$ is also reduced to orthorhombic though the associated distortion is very weak. However, a cubic to orthorhombic structural change at $T_C$ seems to be incompatible with the second-order nature of the magnetic transition \cite{Yamasaki06}. The large mode-splitting ($\Delta\omega/\omega_0$$\approx$10\%) observed in the other two materials implies that spin-phonon coupling is indeed strong in these cases.
\begin{figure}[t!]
\includegraphics[width=3.4in]{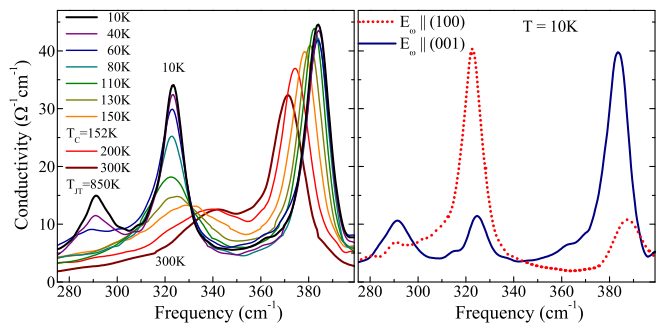}
\caption{(Color online) Optical conductivity of CuCr$_2$O$_4$ in the range of the 2$^{nd}$ phonon mode at various temperatures. The splitting induced by the cooperative Jahn-Teller effect is already present at room temperature while a third, low-energy peak becomes isolated only in the magnetic phase. T$=10$\,K polarized spectra with $\mathbf{E}_{\omega}$ roughly parallel to the tetragonal axis (001) and along (100) are plotted in the right panel.}
\label{fig4}
\end{figure}

Former optical studies on chromite spinels with nonmagnetic A-site ions have found noticeable splitting only for the first two phonon resonances upon the magnetic transition except for ZnCr$_2$S$_4$ where the effect is almost equally manifested in each mode \cite{Rudolf07}. These findings were interpreted in terms of the spin-phonon interaction coming from the spatial dependence of the exchange couplings \cite{Tchernyshyov02,Fennie06,Rudolf07}. Microscopic models explained the large splitting of the two lower-energy T$_{1u}$ modes, showing that the corresponding vibrations mostly affect the Cr-Cr bonds responsible for the nearest neighbor antiferromagnetic interaction. In these highly frustrated magnets the long-range spin order can only be established by the spin Jahn-Teller effect, thus magnetic ordering and structural distortion are necessarily interlocked.

This scenario does not apply for the present spinels. At first, they order at higher temperatures indicating the release of frustration due to the coupling between spins on the pyrochlore and the diamond lattice. With Jahn-Teller active ion on the A site, frustration can be further reduced as manifested in the high T$_C$ of CuCr$_2$O$_4$ which has the strongest tetragonal distortion among the ACr$_2$O$_4$ family. Here we report distinct differences in their vibrational properties. In CuCr$_2$O$_4$ and FeCr$_2$O$_4$, besides the splitting of lower-energy modes we found that the degeneracy is lifted for the higher lying ones, as well. (Hints of splitting in CoCr$_2$O$_4$ again appear in the third and forth modes.) This behavior can be explained by the additional antiferromagnetic A-O-Cr superexchange which is thought to be relevant beyond the antiferromagnetic Cr-Cr direct exchange and the ferromagnetic Cr-O-Cr superexchange only present for nonmagnetic A-site \cite{Menyuk07}. Lattice dynamical calculations on MgTi$_2$O$_4$ \cite{Popovic03} and ZnCr$_2$O$_4$ \cite{Fennie06} predict that T$_{1u}$(3) and T$_{1u}$(4) vibrations have strong contributions from oxygen displacements. Thus, they can effectively modulate the A-O-Cr and Cr-O-Cr exchange path. Since the spins on Cr$_1$ and Cr$_2$ sublattices "align antiparallel" the latter has minor significance and we attribute the magnetically induced lattice distortion to the \emph{spin-phonon coupling corresponding to the A-O-Cr exchange path}.

There is another possible scenario in which the interplay of Jahn-Teller effect and spin-orbit coupling on the A-site ions plays a crucial role. In CoCr$_2$O$_4$ the orbital moment of Co$^{2+}$ ion is quenched and the spin-orbit interaction is weak since it can only mix states separated by the large tetrahedral crystal field. On the other hand, in CuCr$_2$O$_4$ the spin-orbit interaction is more effective as it competes only with the smaller tetragonal component of the crystal field. Therefore, it can generate a larger single-ion anisotropy which may be responsible for the magnetically induced orthorhombic distortion. This is an alternative way to introduce spin-lattice coupling besides the spatial variation of the exchange interactions considered so far in spinels with nonmagnetic A site.

Another peculiarity of these compounds that magnetic ordering is strongly manifested in the lattice dynamics while it is hardly observable in static distortions. Though no structural component of the magnetic transition has been found so far in x-ray and neutron scattering experiments, we could clearly demonstrate the lowering of the crystal symmetry by the analysis of infrared active phonon modes. The only compound, related to this family, where the onset of a tetragonal to orthorhombic distortion was reported at T$_C$ is NiCr$_2$O$_4$ \cite{Ishibashi07}. Even in that case the size of the orthorhombic distortion at low temperatures ($\varepsilon_o$$\approx$$0.18$\%) falls far below the scale of the tetragonal one ($\varepsilon_t$$\approx$$5$\%), the latter of which is similar to that of FeCr$_2$O$_4$ and CuCr$_2$O$_4$ ($\varepsilon_t$$\approx$$2.5$\% and $10$\%, respectively). Phonon modes turn out to be better indicators of the structural transformation, as well, since some of the mode splittings in CuCr$_2$O$_4$ reach $\Delta\omega/\omega_0$$\approx$$40$\%.

In conclusion, we pointed out that in ACr$_2$O$_4$ spinels with magnetic ion on the A-site the lattice symmetry is broken upon their ferrimagnetic transition. The detailed analysis of the infrared active phonon modes showed that in CuCr$_2$O$_4$ and FeCr$_2$O$_4$ the lattice form is lowered to orthorhombic from originally tetragonal in their high-temperature paramagnetic phase. We found that the second magnetic transition of FeCr$_2$O$_4$ also affects the lattice vibrations. In contrast, CoCr$_2$O$_4$ shows tiny deviation from the cubic structure down to T$=10$\,K.

The authors are grateful to K. Penc for enligthening
discussions. This work was supported by a Grant-In-Aid for
Scientific Research, MEXT of Japan, the Hungarian
Research Funds OTKA PD75615, NK72916, Bolyai 00256/08/11, the SFB 484 of the Deutsche Forschungsgemeinschaft and the Bayerische Forschungsstiftung.


\begin{references}
%
\bibitem{Kimura07}T. Kimura, Ann. Rev. Mat. Res. \textbf{37}, 387 (2007).
%
\bibitem{Millis96}A.J. Millis, \prb \textbf{53}, 8434 (1996).
%
\bibitem{Hirota94}K. Hirota et al., \prl \textbf{73}, 736 (1994).
%
\bibitem{Grunerbook}G. Gruner, \textit{Densitiy Waves in Solids}, Addison-Wesley Publishing Company (1994).
%
\bibitem{Dagotto02}M. Uehara et al., Nature \textbf{399}, 560 (1999); Y. Tokura, Rep. Prog. Phys. \textbf{69}, 797 (2006).
%
\bibitem{Ronnow06}H.M. Ronnow et al., Nature \textbf{440}, 1025 (2006).
%
\bibitem{Kezsmarki06}I. K\'ezsm\'arki et al., \prb \textbf{73}, 125122 (2006).
%
\bibitem{Agrait02}N. Agrait et al., \prl \textbf{88}, 216803 (2002).
%
\bibitem{Lee08}S. Lee et al., Nature \textbf{451}, 805 (2008).
%
\bibitem{Tchernyshyov02}O. Tchernyshyov, R. Moessner, and S.L. Sondhi, \prl \textbf{88}, 067203 (2002).
%
\bibitem{Penc04}K. Penc et al., \prl \textbf{93}, 197203 (2004).
%
\bibitem{Roth58}W.L. Roth, Phys. Rev. \textbf{110}, 1333 (1958).
%
\bibitem{Massida99}S. Massidda et al., \prl \textbf{82}, 430 (1999).
%
\bibitem{Fennie06}C.J. Fennie and K.M. Rabe, \prl \textbf{96}, 205505 (2006).
%
\bibitem{Rudolf07}T. Rudolf et al., New J. Phys. \textbf{9}, 76 (2007).
%
\bibitem{Menyuk07}T. A. Kaplan and N. Menyuk, Phil. Mag. \textbf{87}, 3711 (2007).
%
\bibitem{Shirane64}G. Shirane, D.E. Cox, and S.J. Pickart, J. Appl. Phys. \textbf{35}, 954 (1964).
%
\bibitem{Tomiyasu04}K. Tomiyasu, J. Fukunaga, and H. Suzuki, \prb \textbf{70}, 214434 (2004).
%
\bibitem{Yamasaki06}Y. Yamasaki et al., \prl \textbf{96}, 207204 (2006).
%
\bibitem{Prince57}E. Prince, Acta Cryst. \textbf{10}, 554 (1957).
%
\bibitem{Ohgushi08}K. Ohgushi et al., J. Phys. Soc. Jpn. \textbf{77}, 034713 (2008).
%
\bibitem{Tanaka66}M. Tanaka, T. Tokoro, and Y. Aiyama, J. Phys. Soc. Jpn. \textbf{21}, 262 (1966).
%
\bibitem{Kennedy08}B.J. Kennedy and Q. Zhou, J. Solid State Chem. \textbf{181}, 2227 (2008).
%
\bibitem{note}Infrared active E$_u$ modes also derive from the originally silent T$_{2u}$'s due to the tetragonal distortion, though they probably carry negligible oscillator strength.
%
\bibitem{Popovic03}Z.V. Popovic et al., \prb \textbf{68}, 224302 (2003).
%
\bibitem{Ishibashi07}H. Ishibashi and T. Yasumi, J. Magn. Magn. Mater. \textbf{310}, e610 (2007).
\end{references}
\end{document}